\documentclass[conference]{IEEEtran}
\IEEEoverridecommandlockouts
\usepackage{cite}
\usepackage{flushend}
\usepackage{amsmath,amssymb,amsfonts}
\usepackage{graphicx}
\usepackage{textcomp}
\usepackage{xcolor}
\usepackage{tabularx}
\usepackage{lipsum}  
\usepackage[linesnumbered,ruled,vlined]{algorithm2e}
\def\BibTeX{{\rm B\kern-.05em{\sc i\kern-.025em b}\kern-.08em
    T\kern-.1667em\lower.7ex\hbox{E}\kern-.125emX}}
\begin{document}

\title{Energy Efficient WSN: a Cross-layer Graph Signal Processing Solution to Information Redundancy
\thanks{A. Chiumento is funded from the European Union’s Horizon 2020 research and innovation programme under the Marie Skłodowska-Curie grant agreement No 713567 and this material is based upon work supported by the Air force Office of Scientific Research under award number FA9550-18-1-0214 and supported by the Science Foundation Ireland under Grant No. 13/RC/2077.}
}

\author {Alessandro Chiumento, Nicola Marchetti and Irene Macaluso\\
\small
    Connect Centre, Trinity College Dublin, Ireland\\
	\{alessandro.chiumento\}@tcd.ie \\
}

\maketitle

\begin{abstract}
In this work an iterative solution to build a network lifetime-preserving sampling strategy for WSNs is presented. The paper describes the necessary steps to reconstruct a graph from application data. Once the graph structure is obtained, a sampling strategy aimed at finding the smallest number of concurrent sensors needed to reconstruct the data in the unsampled nodes within a specific error bound, is presented. An iterative method then divides the sensor nodes into sets to be sampled sequentially to increase lifetime. Results on a real-life dataset show that the reconstruction RMSE can be easily traded off for a larger number of disjoint sampling sets which improve the network lifetime linearly. 
\end{abstract}


\section{Introduction}

Extending the lifetime of a Wireless Sensor Network (WSN) without compromising its sensing capabilities is a key objective if wireless solutions will ever replace wired ones \cite{murillo2017bluetooth, chiumento2018building}. The great advantage of having untethered sensors is that they can be positioned where most needed or even randomly scattered in an environment. Specific applications in which sensors are difficult to reach or difficult to replace (e.g. sensors embedded in concrete structures \cite{Concrete_sensors} or spread over agricultural areas \cite{Prec_agri}) call then for a careful analysis of lifetime-extending strategies, identifying the key components of energy efficiency in WSNs \cite{RAULT2014104}. Aside from creating more energy efficient sensors, or designing specific radio protocols to enhance energy conservation, a simple yet effective solution is to sense only when and where necessary. The sensing and data forwarding is, in fact, the most energy expensive task performed by the sensor \cite{RAULT2014104}. Unnecessary sample transmissions affect both the lifetime of the sensor node as well as the overall network performance as they generate overhead \cite{7410050}. \textit{Adaptive sampling} techniques adjust the sampling rate of every sensor while ensuring that global performance targets (in terms of overall sensing accuracy) are met. For example, in \cite{Adap_samp_WSN} the authors present a comparative study of different adaptive sampling strategies aimed at increasing the sleep time of each sensor and reducing the sense and transmit intervals. In general, traditional adaptive sampling strategies aim at determining the best sampling intervals of each sensor in order to reduce the duty-cycle, but they do not address the topology of the sensor network and how the sensors might be correlated with one another with regards to the application data being sensed.
In Compressive Sensing (CS), on the other hand, recent work has focused on reducing both the number of sensing nodes as well as compressing the measured data as it gets relayed over the WSN \cite{SRISOOKSAI201237}. Traditional Compressive Sensing techniques do rely on time-domain sparsity or, at the very least, allow for a sparse representation of the signal to be sampled, where a sampled observation is compressed at the sensing node and then the aggregate samples are reconstructed at the sink or receiving node \cite{Compressed_acquisition}. In order to develop a lifetime-preserving scheduling strategy for WSN based on CS techniques, the authors in \cite{CS_scheduling} have presented a probabilistic solution which determines which sensor nodes are correlated (in space and time) and maximise network lifetime by maximising the sleep time of the nodes while keeping coverage high (enough sensors are engaged to cover the sensing area completely). All the above solutions require either sparsity in the time domain or strong time-domain correlation. \\

Current advancements in the field of \textbf{Graph Signal Processing} (GSP) have shown that it is possible to extend traditional signal processing methods for time-varying signals onto irregular structures such as graphs. A graph signal is, in fact, a signal which is sampled over the vertices of a graph rather than on a time line; by incorporating the topology of a graph, it is then possible to include the additional information on how a signal propagates over a graph, across vertices (in a WSN analogy: how sampled measurements are related between sensor nodes) \cite{Puy_2018}. 
As in traditional signal processing, where downsampling a time varying signal means reducing the time-domain samples in order to limit the datarate, in GSP downsampling implies sampling a graph signal over only a subset of nodes. The problem lies thus in reconstructing the complete graph signal for this subset \cite{Rusu_2017, GSP_APPS}. Considerable work has been presented on how to reconstruct a graph signal from few nodes \cite{Puy_2018, Rusu_2017} and on what would be an optimal sampling subset given a limitation on the possible bandwidth of the sampled graph signal \cite{Tanaka_2018, 7366599, Dom_Set_WSN}. 
On the other hand, in WSNs, once sensors are deployed it is often impractical to retrieve them and certainly wasteful to have wireless and battery operated sensors and not use them as they might not be part of the optimal sampling subset. 

In order to guarantee optimal sensing performance and network lifetime, it is then important to take into consideration that:
\begin{itemize}
    \item the sensing performance of the WSN needs to be guaranteed and thus sampling strategies aimed at reducing energy should not impact it;
    \item all sensors nodes should contribute to the sensing and
    \item in order to guarantee robustness, all sensor nodes should be energy depleted at the same time.
\end{itemize}
This work builds on the previous GSP sampling results and presents an iterative sampling strategy which aims to reconstruct the graph signal by dividing the network into disjoint partitions, where sampling each of them will guarantee good signal reconstruction within an arbitrary bound. Each partition might be sampled sequentially, so that the sampling procedure's overall energy cost decreases linearly with the number of partitions.

\section{A Primer on Graph Signal Processing for Signal Sampling}
This section presents a brief overview of the GSP methods used in this work to develop an energy efficient sampling strategy for WSNs. Firstly, the basic notation is given.

Consider an undirected graph $G = (V,E)$ where $V$ represents the set of vertices $V = {1,...,n}$ and $E$ represents the edges connecting the vertices $E = {w_{i,j}}$. $w_{i,j} \in \mathbb{R}_{+}$ is the weight associated to the edge between vertices $i$ and $j$. The adjacency matrix $\textbf{W}$ can then be defined as a matrix containing the weights of the edges between vertices, with elements $\textbf{W}_{i,j} = w_{i,j}$; if an edge is missing the associated weight will be zero. The degree matrix $\textbf{D}$ indicates the relative importance of each vertex with respect to all the other ones and is defined as $\textbf{D} = diag(d_{1}, ... d_{n})$ where $d_{i}$ is the number of edges attached to each vertex and is the degree of vertex $i$. Using the degree matrix $\textbf{D}$ and the adjacency matrix $\textbf{W}$ one can define the graph \textit{Laplacian matrix} as $\textbf{L} = \textbf{D} - \textbf{W} $.

According to the spectral decomposition theorem, there exists a matrix $\textbf{U}$ such that 
\begin{equation}
\mathbf{L} = \mathbf{U}\mathbf{\Lambda}\mathbf{U}^T.    
\label{Spec_decomp}
\end{equation}
where the columns of $\mathbf{U} = [\mathbf{u_1} \mathbf{u_2} ... \mathbf{u}_N]$ are the eigenvectors of $\mathbf{L}$ and $\mathbf{\Lambda} = diag(\lambda_{1},\lambda_{2},...,\lambda_{N},)$ is a diagonal matrix containing the N eigenvalues $\lambda_{i}$ associated with the eigenvectors $\mathbf{u_i}$ in $\mathbf{U}$.\\
 $\mathbf{U}^T$ is the Graph Fourier Transform (GFT), which contains information on the variability of signals over the graph in a similar way as the Fourier transform does for time-domain signals \cite{GSP_Brain2018}.
If one defines a \textit{graph signal} $\mathbf{x}$ as the signal sampled over all the vertices of the graph, then the GFT of  $\mathbf{x}$ can be defined as
\begin{equation}
\mathbf{\Tilde{x}}(i) = \sum_{k=1}^{N} \mathbf{x}(k) \mathbf{u}_{i}(k) = \mathbf{U}^{T}\mathbf{x}
\label{GFT}
\end{equation}
The eigenvalues of the Laplacian  are then the analogous of frequencies and the eigenvectors are the Fourier basis.

\subsection{Building a graph from measurements}
Given a set of measurements over the sensor nodes, the first step is to determine the graph structure that better describes the relationships between the sensors themselves. In \cite{pmlr-v51-kalofolias16}, the authors have shown a way to learn the graph structure under smoothness assumption of the graph signals.  This entails that, given $T$ measurements on each node of the network, the set of measurements can be defined as a $T \times N$ matrix $\mathbf{X} = [\mathbf{x}_1 \mathbf{x}_2 ... \mathbf{x}_N]$ where each vector $\mathbf{x}_i$ contains the $T$ samples measured over each node $i$. The columns of $\mathbf{X}$ are thus the graph signal.
The smoothness assumption in \cite{pmlr-v51-kalofolias16} requires that the differences between signals in connected nodes is small. The distance between signals of well connected nodes can be quantified as \cite{pmlr-v51-kalofolias16}:
\begin{equation}
\frac{1}{2} \sum_{i,j} w_{i,j} || \mathbf{x}_{i} - \mathbf{x}_{j} ||
\label{Sig_dist}
\end{equation}
which in matrix form becomes:
\begin{equation}
tr(\mathbf{X}^{T} \mathbf{L} \mathbf{X}) .
\label{Sig_dist_mat}
\end{equation}
There is then a graph Laplacian $ \mathbf{L}$ which will minimise the distance between signals of connected nodes. The graph construction process is thus the search for this Laplacian by performing the following optimisation:
\begin{equation}
\min_{L}      tr(\mathbf{X}^{T} \mathbf{L} \mathbf{X}) + f(\mathbf{L}),
\label{minim_L}
\end{equation}
where $f(\mathbf{L})$ is a function which prevents $\mathbf{L}$ from being trivial, such as containing only zeros, and can impose further structure using prior information on the graph \cite{pmlr-v51-kalofolias16}. Since $\mathbf{X}$ is known, by solving (\ref{minim_L}) it is possible to directly learn the graph topology behind the data \cite{pmlr-v51-kalofolias16}.

\subsection{Reconstructing a signal from a subset of nodes}
It is now possible to make use of the discovered graph Laplacian to determine the graph signal's missing values. 

The well known \textit{Total Variation} principle states that, if the graph signal is smooth, there is a limit to the variation that a signal can have going from a vertex $i$ to its adjacent vertices \cite{7746675}. Similarly then to Equation (\ref{minim_L}) the \textit{Total Variation} of a graph can be expressed as:
\begin{equation}
TV_{G}(\mathbf{x}) = \sum_{i=1}^{N} |x(i) - \hat{x}(i)| =  || \mathbf{x} - \mathbf{W}\mathbf{x} ||_{1},
\label{TV}
\end{equation}
in which $\hat{x}(i)$ is the shifted signal at the neighbours of vertex $i$. In the quadratic form, (\ref{TV}) becomes:
\begin{equation}
TV^{q}_{G}(\mathbf{x}) = \frac{1}{2}||\mathbf{x} - \mathbf{W}\mathbf{x} ||_{2}^{2},
\label{TV2}
\end{equation}
Given a graph signal in the form $\mathbf{x} = \begin{bmatrix}     \mathbf{x}_{M}\\ \mathbf{x}_{U} \end{bmatrix}$ in which $\mathbf{x}_{M}$ represents the known signal sampled over $M$ nodes and $\mathbf{x}_{U}$ is the missing signal which needs to be reconstructed, it is possible then to recover the missing signal by solving the following unconstrained optimisation \cite{Chen_GSP_2014}:
\begin{equation}
\mathbf{x}^{*} = \mathbf{argmin} \frac{1}{2}||\hat{\mathbf{x}}_{M} - \mathbf{x}_{M} ||_{2}^{2} + \eta TV^{q}_{G}(\mathbf{x}).
\label{min_TVG}
\end{equation}
In Equation (\ref{min_TVG}), $\eta$ is a tuning parameter which controls the impact of the smoothness criterion on the minimisation process. A large $\eta$ places more weight on the smoothness of the solution while a smaller $\eta$ puts more emphasis on fitting a solution onto the known measurements. As Equation (\ref{min_TVG}) is a convex quadratic problem, it has a closed-form solution, and in turn this means that $\eta$ can be tuned efficiently based on actual graph signals.

\subsection{Optimal Sampling Strategy}
The authors in \cite{Puy_2018} have shown that there is a random sampling strategy that guarantees a bound on the reconstruction error in noisy signals, based on sampling the nodes where the signal energy is most concentrated.
The method works by determining the number $n$ of vertices that need to be sampled to enable correct reconstruction. Let $\mathbf{m}$ be a sampling vector (in which $\mathbf{m}_{i} = 1$ if vertex $i$ is sampled and zero otherwise), then for any $\delta \in ]0,1[$:
\begin{equation}
(1 - \delta)||\mathbf{x}||_{2}^{2} \leq \frac{N}{n}||\mathbf{m}\mathbf{x}||_{2}^{2} \leq (1 + \delta)||\mathbf{x}||_{2}^{2}.
\label{random_sampling}
\end{equation}

Equation (\ref{random_sampling}) (Theorem 2.2 in \cite{Puy_2018}) is a sufficient condition for making sure that $\mathbf{m}$ satisfies the bandlimitness of the graph signal and indicates that sampling $O(k \log k)$ vertices is enough to recover the complete signal, where $k$ is the index of the highest spectral component in the graph signal. This entails that, given a chosen $\delta$ there exists an $\mathbf{m}$, which represents the sampling mask, allowing to reconstruct the signal correctly.

\section{An Iterative Graph Partitioning Strategy}
\label{sec:iter}
The solution proposed in this work combines the previous results into a greedy iterative heuristic search, aiming at sampling the lowest possible number of nodes while keeping the reconstruction error as small as possible. The reconstruction error is thus defined by using the \textit{Root Mean Square Error}:
\begin{equation}
RMSE = \sqrt{\frac{\sum_{i=1}^{N} \hat{x}(i) - x(i)}{N} } ,
\label{MSE}
\end{equation}
in which $\hat{x}(i)$ is the reconstructed graph signal at vertex $i$.
First the graph is constructed from successive sampled measurements by solving problem (\ref{minim_L}). By solving problem (\ref{random_sampling}), a sampling mask $\mathbf{m}$ which satisfies the bandlimitness of the graph signal given the signal measurements and the graph topology is found. The sampling mask derived from (\ref{random_sampling}) provides which nodes contain the highest amount of information. The nodes are then placed in descending order to be sampled using disjoint sets containing the fewest number of nodes per set while still respecting the imposed RMSE reconstruction threshold.

The proposed solution selects the best node according to (\ref{random_sampling}) and reconstructs the signal using (\ref{min_TVG}). If the RMSE computed according to (\ref{MSE}) is below the given threshold then the selected node becomes a set. Otherwise, the next node in the list is added in the sampling set and the process continues until the RMSE is below threshold; in which case the set is complete. A new sampling set begins by selecting the next best node. The process is continued until all the nodes are sampled. Algorithm \ref{algo_sample} shows the whole procedure.
\small{
\begin{algorithm}
\SetAlgoLined
\KwResult{The variable $\mathbf{S}$ contains the disjoint sampling sets which satisfy the reconstruction error condition. }
 initialisation: RMSE = $\infty$, Sampled nodes = [], Choose threshold $\epsilon$\;
 Receive $\mathbf{X}$ signal from nodes\;
 Build graph according to (\ref{minim_L})\;
 Determine sampling distribution by using (\ref{random_sampling})\;
 Order nodes in descending order of sampling importance in list $L$\;
\While{Sampled nodes $<$ total nodes $N$}{
 Sampling mask $\mathbf{m} = []$\;
    \While{RMSE $> \epsilon$}{
        Select first node in $L$ and add to sampling mask $\mathbf{m}$\;
        Reconstruct graph signal ${\hat{\mathbf{X}}}$ using (\ref{min_TVG})\;
        Compute RMSE according to (\ref{MSE})\;
        Remove sampled node from list $L$\; 
        }
    Store sampling mask $\mathbf{m}$ in set $\mathbf{S}$\;
    Add selected nodes to Sampled node list;
    }
\Return{} $\mathbf{S} $ 
 \caption{Iterative Sampling Strategy}
 \label{algo_sample}
 
\end{algorithm}
} 
\begin{figure}[t]
	\centering	
	\includegraphics[width=0.40\textwidth]{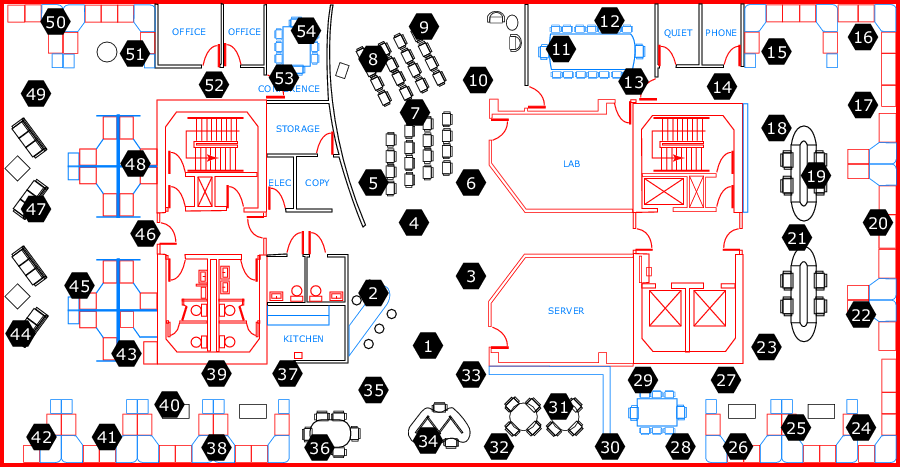} 
	\caption{Sensor network used in\cite{Intel_Data} to collect the measurements. The sensors are indicated as black dots and they all communicate with a central controller.}
	\label{iot}
\end{figure}
\section{The Wireless Sensor Network}
The  dataset  used  in  this  paper is composed by temperature measurements collected by distributed sensors in an indoor environment; the dataset is openly available and was collected by the "Intel Research Berkeley Lab" \cite{Intel_Data}. The sensor network is composed by 54 Mica2Dot wireless nodes positioned in an area of 1200 $m^{2}$; each sensor measures local temperature, sampling the environment every 31 seconds and sends the measured value to a unique gateway.
 
As the communication between the sensors and the gateway is wireless, packet losses are present due to fluctuating channel quality and collisions. This entails that not all measurements are received from all nodes at the same time and thus the complete information over all the sensors needs to be collected during multiple time intervals.

\section{Results}
\begin{figure}[h]
	\centering	
	\includegraphics[width=0.50\textwidth]{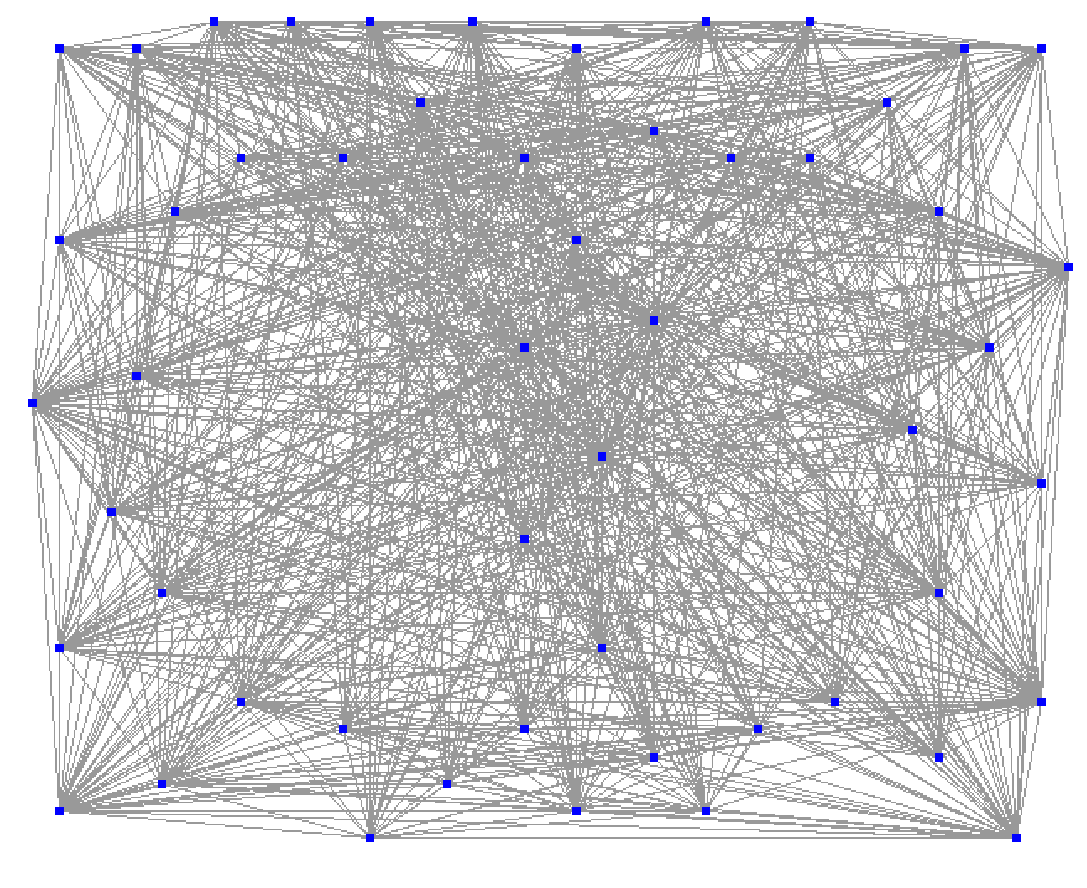} 
	\caption{Final graph built from consecutive measurements. It is visible that all the nodes are well connected and thus the measurements present strong correlation in the vertex domain, that is to say the graph signal is smooth. }
\label{fig:graph}
\end{figure}

\begin{figure}[h]
	\centering	
	\includegraphics[width=0.50\textwidth]{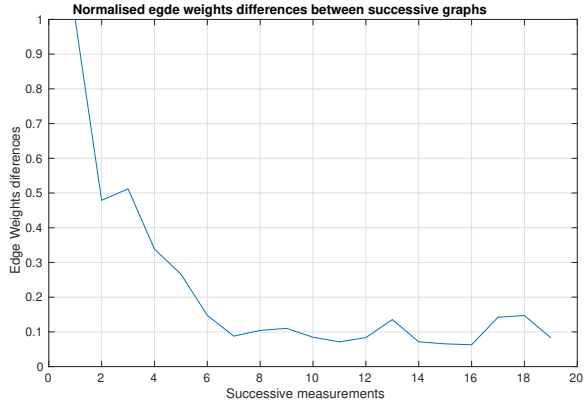} 
	\caption{Convergence time of the Laplacian as a function of the received graph signal. As new measurements arrive, new nodes partake in the graph formation until the structure is stable.}
\label{fig:graph_built}
\end{figure}

As described in Section \ref{sec:iter} the algorithm first builds a graph from successive collected measurements, Figure \ref{fig:graph} shows the graph built from 10 consecutive measurements. As some packets containing sampled data get lost due to poor channel quality or collisions, the graph is built from consecutive measurements until all nodes are sampled and the resulting graph Laplacian does not change anymore. Figure \ref{fig:graph_built} shows the convergence time until the Laplacian of the discovered graph has stabilised in function of the received measurements.

\begin{figure}[h]
	\centering	
	\includegraphics[width=0.50\textwidth]{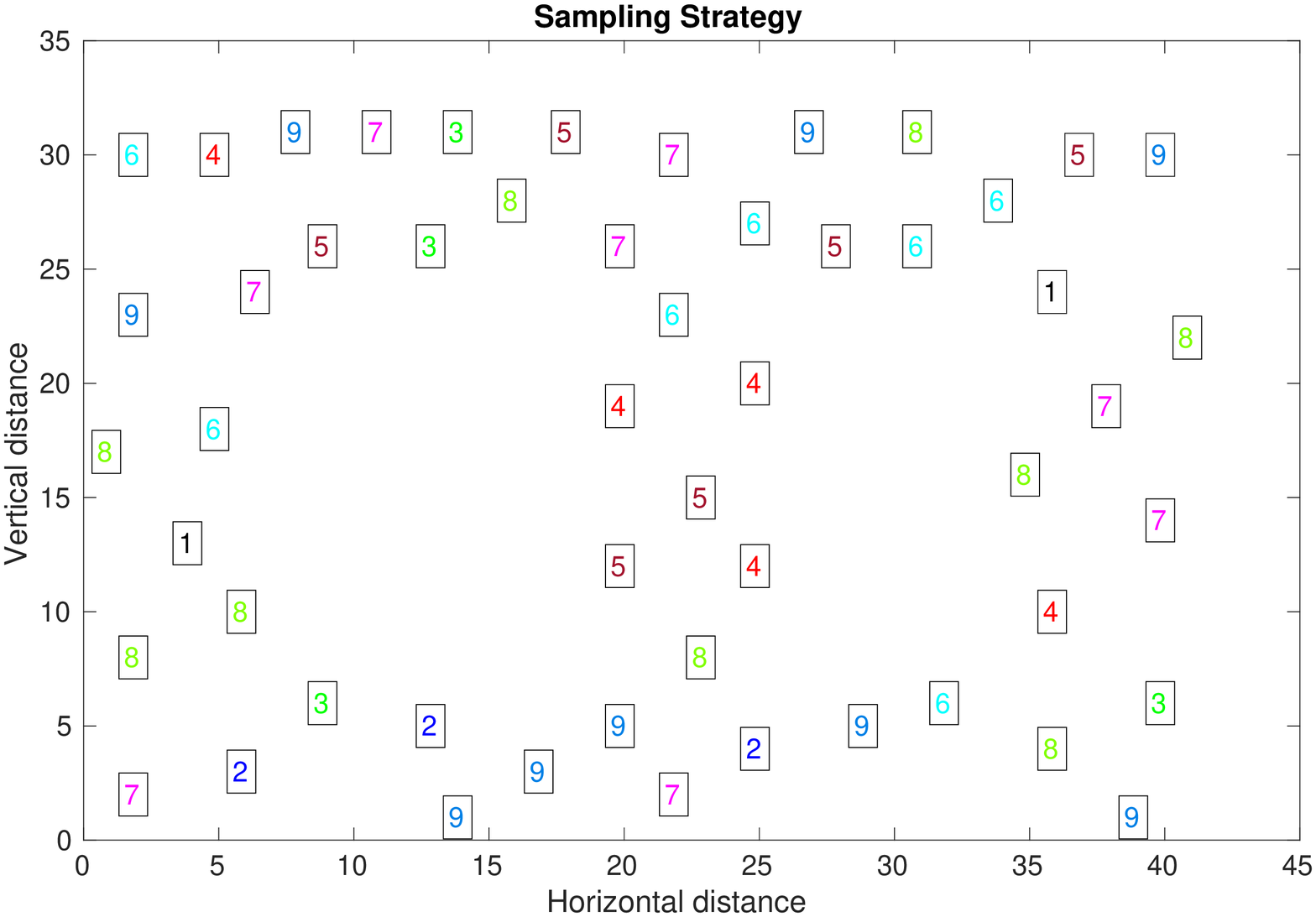} 	
	\caption{The sensors are split into 9 disjoint sampling sets when the Threshold  $\epsilon = 0.3$ is selected. The signals sampled from each set allows a reconstruction with an RMSE within the selected threshold.}
\label{fig:sets_when_mse_fixed}
\end{figure}

Once the graph is complete, Algorithm \ref{algo_sample} determines the disjoint sampling sets iteratively. Figure \ref{fig:sets_when_mse_fixed} presents the sampling sets results when the MSE threshold is fixed at $\epsilon = 0.3$. The 54 sensors are then divided into 9 disjoint sets, each of which is able to reconstruct the signal within the given RMSE threshold.

\begin{figure}[h]
	\centering	
	\includegraphics[width=0.50\textwidth]{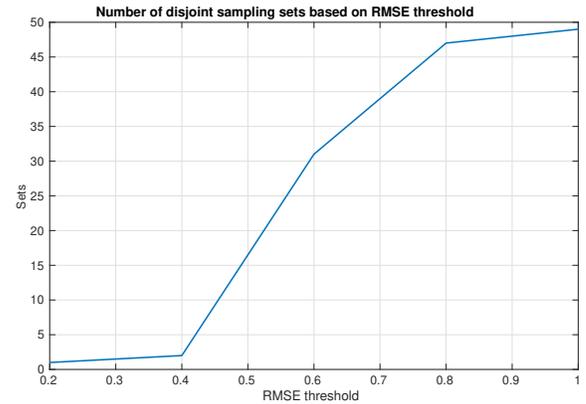} 	
	\caption{The sensors are split into different sets of varying size depending on the RMSE threshold set at the sampling selection strategy. }
\label{fig:sets_vs_MSE}
\end{figure}

To showcase the impact of the threshold selection in the set creation and thus the overall number of concurrent sensors engaged, Figure \ref{fig:sets_vs_MSE} presents the number of disjoint sets as a function of the RMSE threshold $\epsilon$. As the the threshold increases, a lower resolution is necessary and thus a smaller number of concurrent sensors is selected in the sampling mask $M$ to reconstruct the signal. This implies that the number of sets increases as fewer sensors are engaged at the same time. The number of sets is then monotonically dependent on the threshold but not necessarily linear, in fact, in the studied graph the curve appears to have a sigmoid behaviour as small values of the RMSE threshold require the engagement of large number of nodes to bound the error and the number of sets grows slowly. Very large values of the RMSE threshold, on the other hand, slow down the creation of new sets as the size of each set approaches the single sensor.

While the RMSE threshold bounds the maximum error allowed per set, the actual RMSE measured per set might be lower than the imposed threshold. Since the graph signal reconstruction is highly dependent on the inferred graph topology, the proposed greedy approach combines nodes so that the worst case performance of the discovered sets is within the chosen bound.

\begin{figure}[h]
	\centering	
	\includegraphics[width=0.50\textwidth]{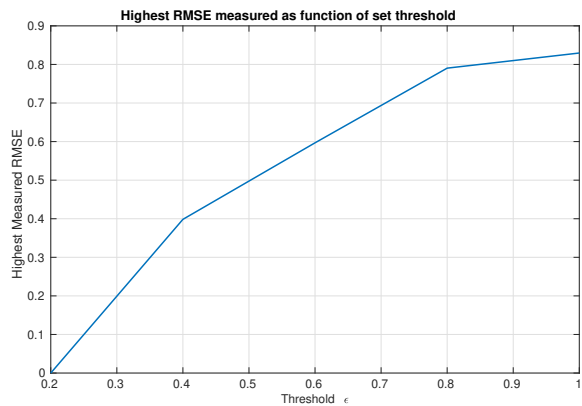}
	\caption{Highest measured RMSE as a function of the number of the selected RMSE threshold. The proposed method provides sets with RMSE consistently lower than the threshold. }
\label{fig:sets_vs_MAXMSE}
\end{figure}

Figure \ref{fig:sets_vs_MAXMSE} shows the highest RMSE measured as a function of the chosen threshold. The behaviour has the same monotonic nature as in Figure \ref{fig:sets_vs_MSE}, but the values are consistently lower than the thresholds. This signifies that, although effective, the proposed method might be improved by selecting nodes using an optimisation process in which the given RMSE threshold is met exactly in very set. This would provide a larger number of sets at the cost of increased complexity.

The relation between number of sets and energy efficiency of the system is evident from Table \ref{table: EnergyVSsets}. The table presents the duty cycle per sensor as a function of the total number of sets. As more sets are selected, fewer concurrent sensors need to be used, and this decreases the amount of energy per sensor linearly with the the number of sets. Depending on the application requirements, it is then possible to trade reconstruction error for energy expenditure. Following the results above, as the RMSE threshold $\epsilon$ increases, the WSN can be divided into an increasing number of disjoint sets. By definition, activating each set of sensors sequentially would enable the same sensor network to live longer while keeping the error bound within the chosen threshold.

 \begin{table}[h]
    \caption{Sensor duty cycle as function of number of sets.}
    \begin{tabularx}{\columnwidth}{||X||c c c c c c||}
        \hline
        Number of Sets & 1   &  2  &  21 &   44  &  49  &  51 \\
        \hline
        Sensor Duty Cycle [\%]  & 100  & 50  &  4.76  &  2.27 & 2.041 & 1.96 \\
  \hline \hline
    \end{tabularx}
    \label{table: EnergyVSsets}
\end{table}

\section{Conclusions}
This paper describes a novel graph sampling method designed to keep the signal reconstruction error bound within a threshold while increasing a sensor network's lifetime. The proposed iterative solution is simple and makes use of all the nodes in the network, in order to ensure even utilisation and maximise lifetime by choosing the highest number of disjoint sampling sets which guarantees reconstruction performances. The provided graph signal processing framework enables the reader to adapt the proposed solution to various graph sampling problems in which the reconstruction process might need to be adjusted for specific applications such as distributed learning over graphs for WSNs, traffic analysis, or biological networks.

\bibliographystyle{IEEEtran}
\bibliography{biblio}

\end{document}